\documentclass[prb,twocolumn,showpacs,superscriptaddress, amsmath,amssymb]{revtex4-2}

\usepackage{dcolumn}
\usepackage{bm}
\usepackage[pdftex]{graphicx}
\usepackage{xcolor}

\newcommand{\vect}[1]{\mathbf{#1}}

\newcommand{\LiCu}{$\rm LiCu_3O_3$}

\begin{document}
\title{Magnetic properties of LiCu$_3$O$_3$ -- quasi-two-dimensional antiferromagnet on depleted square lattice}

\author{A.~A.~Bush}
\affiliation{MIREA - Russian Technological University,
pr. Vernadskogo 78, Moscow, 119454 Russia}

\author{S.~K.~Gotovko}
\affiliation{P.L. Kapitza Institute for Physical Problems, RAS, Moscow 119334, Russia}
\affiliation{National Research University Higher School of Economics, 101000 Moscow, Russia}

\author{V.~Yu.~Ivanov}
\affiliation{Prokhorov General Physics Institute, RAS, 119991 Moscow, Russia}

\author{V.~I.~Kozlov}
\affiliation{MIREA - Russian Technological University,
pr. Vernadskogo 78, Moscow, 119454 Russia}
\affiliation{P.L. Kapitza Institute for Physical Problems, RAS, Moscow 119334, Russia}

\author{E.~G.~Nikolaev}
\email{nikolaev@kapitza.ras.ru}
\affiliation{P.L. Kapitza Institute for Physical Problems, RAS, Moscow 119334, Russia}

\author{L.~E.~Svistov}
\email{svistov@kapitza.ras.ru}
\affiliation{P.L. Kapitza Institute for Physical Problems, RAS, Moscow 119334, Russia}

\date{\today}
\begin{abstract}
\LiCu\ is a novel 2D S=1/2 antiferromagnet with randomly depleted square lattice. The crystal structure contains two types of square planes with different Cu$^{2+}\rightarrow$ Li$^{+}$ substitution rates (20\% and 40\%).  
$^7$Li~NMR and magnetization measurements performed on single crystals of \LiCu\ revealed the occurrence of magnetic order at $T_{c1}=123$~K and the change of the magnetic state at $T_{c2}\approx30$~K. The high temperature transition can be attributed to establishment of magnetic order in planes with higher concentration of magnetic ions and the low temperature transition -- to the magnetic ordering in planes with lower concentration of magnetic ions. Broad continuous NMR spectra below $T_{c1}$ reflect a continuous distribution of values or directions of magnetic moments in \LiCu\ typical for spiral, spin-modulated magnetic structures or structures with frozen disorder. Magnetization measurements revealed a spin-flop transition which indicates weak uniaxial anisotropy of the spin structure. 
Relatively small value of magnetic susceptibility at all orientations of applied magnetic field shows that magnetic structure is rigid,
since the estimated value of saturation field derived from differential susceptibility measured at $\mu_0 H=5$~T  is $\mu_0H_{sat} \approx$200~T. 
\end{abstract}
\maketitle

\section{Introduction} \label{Intro}

Quasi-two-dimensional antiferromagnets on square lattice are of a great interest as experimental 
implementations of models in which the existence of various exotic magnetic states is predicted. Magnetic behavior of such objects is stipulated by the value of spins, hierarchy of exchange interactions of ions in sites of square lattice, and interaction with crystallographic surrounding. Nowadays, a big amount of such magnets with weakly bound crystallographic planes is known~(see, for example, Refs.~\cite{Melzi_2000,Melzi_2001,Shirane_1987, Endoh_1988,Vajk_2002}). Particular interest is shown in strongly quantum case of spin S=1/2 as quantum fluctuations play a crucial role in the choice of established magnetic states. 

Naturally, 
various irregularities of square lattice significantly influence magnetic states. Particularly, special attention is drawn to substitution of magnetic ions with non-magnetic impurities. 
Small dilution leads to change of boundaries of magnetic phase diagram, whereas the strong dilution can cause an occurrence of new exotic magnetic states. Such dilution decreases the spin stiffness, which can lead to an occurrence of microscopically non-uniform phases like frozen disorder or spin glass state.

Models of diluted 2D magnets on square lattice were considered in Refs.~\cite{Kato_2000,Sandvik_2002, Chernyshov_2002, Chernyshov_2013}. Experimentally, square lattice quantum antiferromagnets with different substitution degrees were studied in the systems La$_2$Cu$_{1-x}$(Zn,Mg)$_x$O$_4$~\cite{Vajk_2002,Caretta_2011} and Li$_2$V$_{1-x}$Ti$_x$OSiO$_4$~\cite{Pappinutto_2005}. Here we present a novel example of highly diluted square-lattice quantum antiferromagnetic (SLQAF) material \LiCu.

\LiCu\ is a representative of a cuprate with mixed valency of copper ions: the number of magnetic Cu$^{2+}$~(S=1/2) ions is twice as much as the number of nonmagnetic Cu$^{+}$ ions. The X-ray diffraction experiments on microcrystals of \LiCu\ \cite{Hibble_1990} show that the crystal structure of \LiCu\ consists of four alternating square planes stacked along $C_4$-axis: one square plane consists of Cu$^{+}$ ions and other three square planes consist of (Li$^{+}$,Cu$^{2+}$)O$^{2-}$ complexes. The sites of these three square planes are occupied by the nonmagnetic Li$^+$ ions and magnetic Cu$^{2+}$ ions statistically with proportion 1 to 2 to ensure electrical neutrality. Single-phase samples of  \LiCu\ crystals of millimeter size  were grown recently~\cite{Bush_2019}. The crystallographic structure of \LiCu\ allows to consider the magnetic material as an example of a highly diluted  quasi-two-dimensional $S=1/2$ magnet on a square lattice. Here we discuss the study of magnetic state of \LiCu\  with magnetometry and NMR techniques. Two magnetic transitions were observed at $T_{c1}=123$~K and $T_{c2}\approx 30$~K. In our report we suggest magnetic state of \LiCu\ which can qualitatively describe the set of obtained experimental data.

\section{Crystal structure of L\lowercase{i}C\lowercase{u}$_3$O$_3$}

\begin{figure}[tb]
\includegraphics[width=0.85\columnwidth]{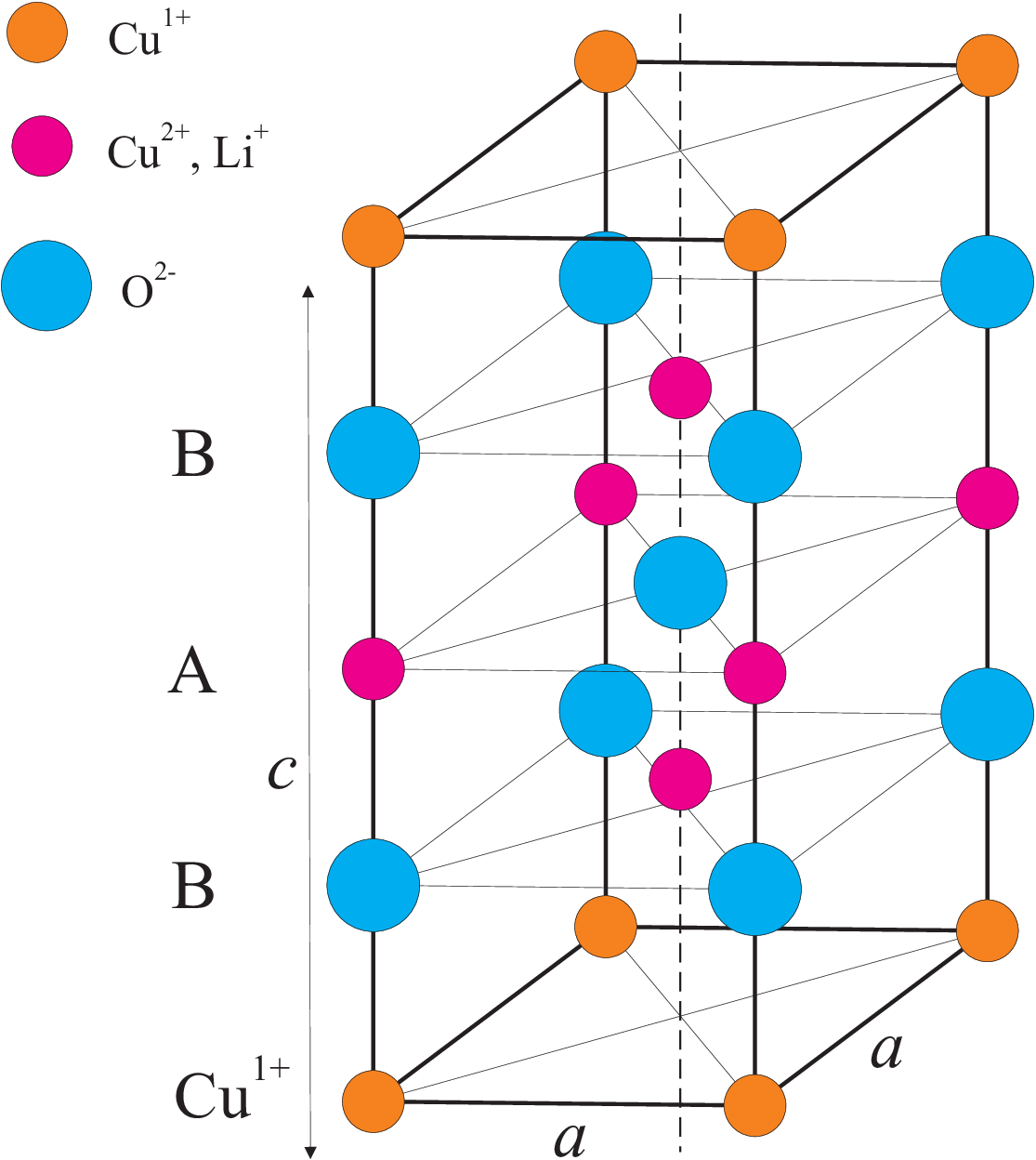}
\vspace{0 mm}
\caption{(color online) Crystal structure of \LiCu.
The smaller orange spheres mark the positions of the Cu$^{+}$ nonmagnetic ions, while the smaller magenta spheres in B,A,B square planes mark the positions of randomly distributed ions of nonmagnetic Li$^{+}$ and magnetic Cu$^{2+}$. The larger blue spheres are at the positions of the O$^{2-}$ ions.}
\label{fig1_crystal}
\end{figure}

\begin{figure}[t]
\includegraphics[width=0.5\columnwidth]{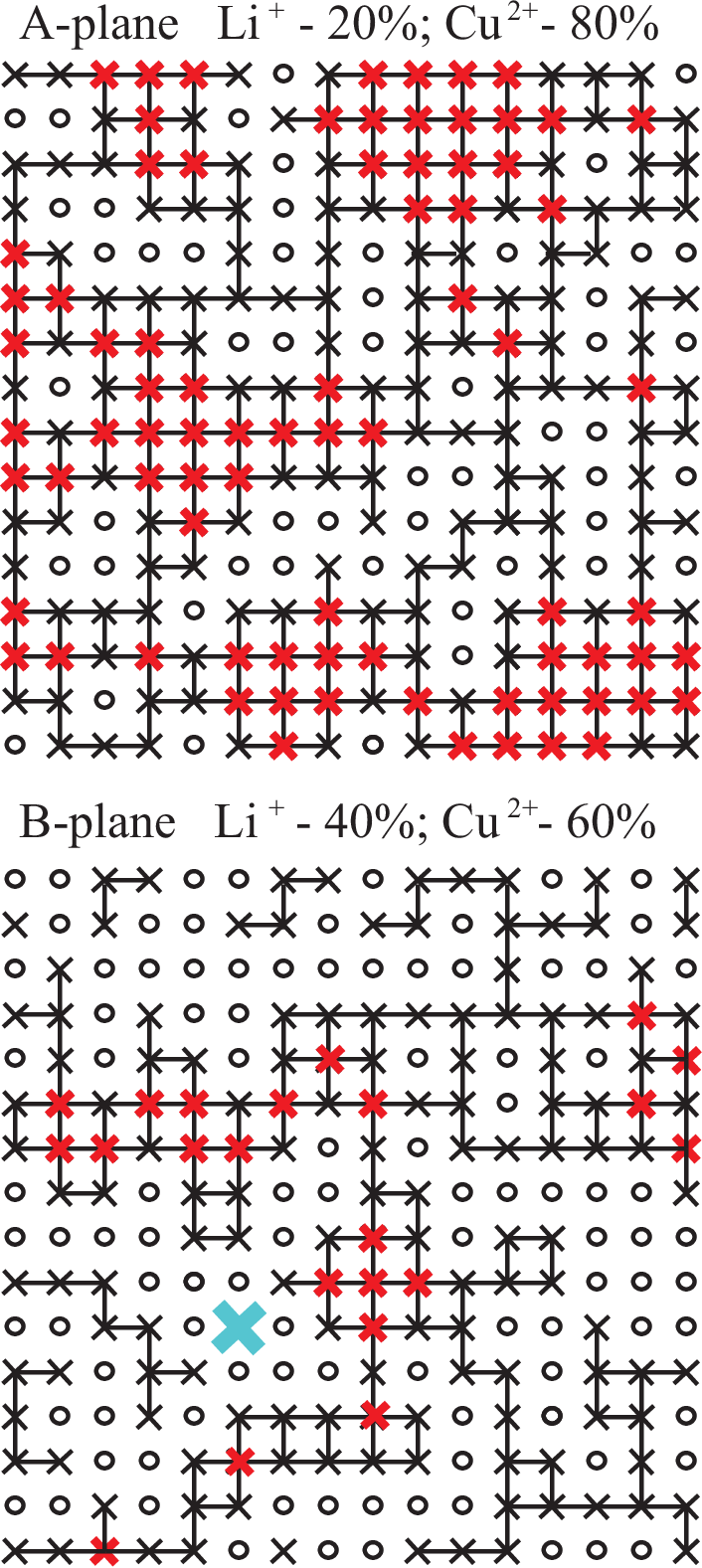}
\caption{(color online) Fragments of simulated square planes A and B with random occupation of crystal nodes of a regular square lattice by magnetic ions Cu$^{2+}$ (crosses) and nonmagnetic Li$^{+}$ (circles).  
The bonds between nearest magnetic ions are shown with lines. Ions which have all four nearest magnetic neighbors are highlighted with bold red crosses. The ion which has no in-plane magnetic neighbors is highlighted with bold blue cross. }
\label{fig2_schema}
\end{figure}

At room temperature, the crystal structure of \LiCu\ belongs to the space group $P_{4}/mmm$. The lattice parameters of \LiCu\ are $a=2.810$~\AA\ and $c=8.889$~\AA~\cite{Hibble_1990}.
The crystal structure consists of alternating layers of Cu$^{+}$ ions and (Li$^{+}$,Cu$^{2+}$)O$^{2-}$ complexes normal to the fourfold axis $C_4$.
Within the layers, the crystal nodes form regular square lattices.
Fig.~\ref{fig1_crystal} shows the schema of crystallographic cell of \LiCu\ drawn in accordance with the X-ray and neutron diffraction experiments given in Refs.~\cite{Hibble_1990,Berger_1992,Bush_2019}.
The planes of nonmagnetic Cu$^{+}$ ions are separated by three (Li$^{+}$,Cu$^{2+}$)O$^{2-}$ square planes. The positions of Cu$^{+}$ and O$^{2-}$ ions are shown in Fig.1 with orange and blue circles respectively. The positions of three intermediate planes colored with magenta are occupied by magnetic ions Cu$^{2+}$ and nonmagnetic ions Li$^{+}$ statistically, forming the solid solution in these planes. The proportion of magnetic copper ions to lithium ions in central plane is $8:2$, whereas for two outer planes the proportion is $6:4$. In the following we will denote the inner planes as A-planes and outer planes as B-planes.

The magnetic properties of \LiCu\ crystals had not been studied before. We can do some comments concerning the expected magnetic structure of \LiCu\ from considering its crystal structure.  First, according to crystallography, the B-A-B triads of magnetic planes are separated by planes of nonmagnetic ions Cu$^{+}$. This fact can suggest their magnetic quasi-two-dimensionality.  The quasi-2D magnetic behavior observed in related compound with mixed valency of copper LiCu$_2$O$_2$~\cite{Matsuda_2005} supports this assumption: in the case of LiCu$_2$O$_2$, nonmagnetic Cu$^{+}$ planes separate pairs of magnetic plains. The second very unusual and attractive feature of the crystal structure is random substitution of magnetic copper ions by nonmagnetic lithium ions:  20\%~for A-plane and 40\% for B-planes. These are very high substitution levels, especially for 2D-systems. The concentration of magnetic copper ions in B-planes is close to percolation threshold for 2D square lattice~\cite{Ziman}. For visualization, we generated two sets of square lattices with rates corresponding to A and B planes presented in Fig.~\ref{fig2_schema}. Copper ions are marked with crosses, and lithium ions are marked with circles. Lines show bonds between nearest magnetic ions. It can be seen that there is a very small number of magnetic ions with full set of in-plane magnetic neighbors:~0.8$^5=0.32$ and 0.6$^5=0.078$ for A- and B- planes, respectively. Such ions are colored with bold red crosses. It can be seen that the areas with regular bonds are coupled with net of ``bridges'' of ions with incomplete sets of exchange bonds. Not all magnetic copper ions have magnetic nearest neighbors: an  example of such isolated ion is marked in the Fig.~2b with bold blue cross. Concluding, we can expect, that magnetic state of \LiCu\ is defined by the random distribution of in- and out-of-square planes exchange interactions.

\section{Sample preparation and experimental details}
\LiCu\ single crystals were grown from high-temperature solutions of Li$_2$CO$_3$ and CuO  in air atmosphere. 

Crystal growth was conducted with flux method. Homogenized batch of 0.125$\cdot$Li$_2$CO$_3\cdot 0.875\cdot$CuO was heated for three hours up to $T=1270^\circ$~C, which is higher than the melting point; then the melt was kept at this temperature for an hour; then it was cooled down to $T=1090^\circ$~C in 30 minutes; after that, it was cooled down to $T=915^\circ$~C in 50 hours; after keeping it at this temperature for 10 hours, the crystallized melt was rapidly cooled down to room temperature~\cite{Bush_2019}.

Typical dimensions of a crystal were $5 \times 5 \times 1$~mm$^3$ with the smallest side aligned with $C_4$ axis of the crystal.
The developed planes of the plate were mirror-like smooth and had the shapes close to a rectangular with sides directed along [110] and [1-10] axes of the crystal. Unit cell parameters were in agreement with the previously reported~\cite{Bush_2019}. The crystals of \LiCu\ were stable in air and did not need any special precautions usually required for some other lithium and sodium cuprates.

The magnetization properties of \LiCu\ were studied with use of commercial magnetometer MPMS-5XL (Quantum Design).

$^{7}$Li nuclei (nuclear spin $I=3/2$, gyromagnetic ratio $\gamma=16.5471$~MHz/T) were probed using the pulsed NMR technique.
The spectra were obtained by summing fast Fourier transforms (FFT) of spin-echo signals as the frequency was swept through the resonance line.
NMR spin echoes were obtained using $\tau_p - \tau_D - \tau_p$ pulse sequences, where the pulse lengths $\tau_p$ were 4-6~$\mu$s and the times between pulses $\tau_D$ were 30-60~$\mu$s. The measurements were carried out in the temperature range $4.2 \leq T \leq 210$~K, temperature stability was better than 0.1~K.

\section{Experimental results}
\subsection{Magnetization measurements}

\begin{figure}[tb]
\includegraphics[width=1\columnwidth]{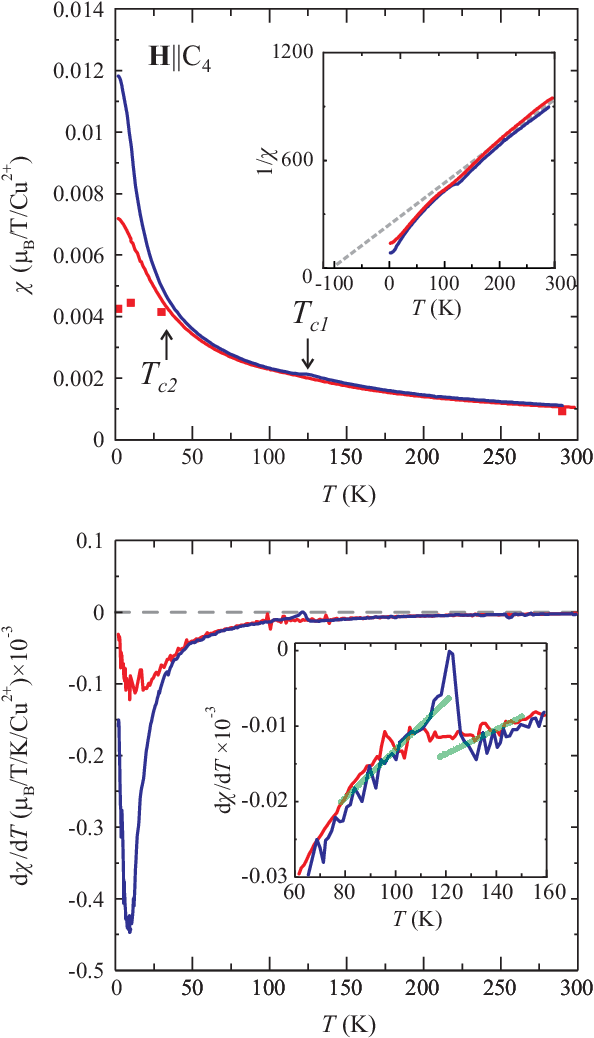}
\caption{(color online) Top panel: Temperature dependence of magnetic susceptibility $\chi(T)$ of \LiCu measured at $\mu_0H=1$~T (blue line) and $\mu_0H=5$~T (red line), $\vect{H}\parallel C_4$. Red squares show temperature dependence of the value of differential susceptibility obtained from $M(H)$ measurements at $\mu_0 H=5$~T (see Fig.~\ref{fig6_M(H1)}). Inverse susceptibilities $1/\chi(T)$ are shown in the inset to the Figure 
with respective colors.
Bottom panel: Temperature derivatives of $\chi(T)$ from top panel. In the inset an expanded area in the vicinity of the singularity observed at $T\approx120$~K is presented. Green solid lines mark the smoothed slopes of $\chi(T)$ at both sides of the critical point.}
\label{fig3_M(T1)}
\end{figure}

\begin{figure}[tb]
\includegraphics[width=1\columnwidth]{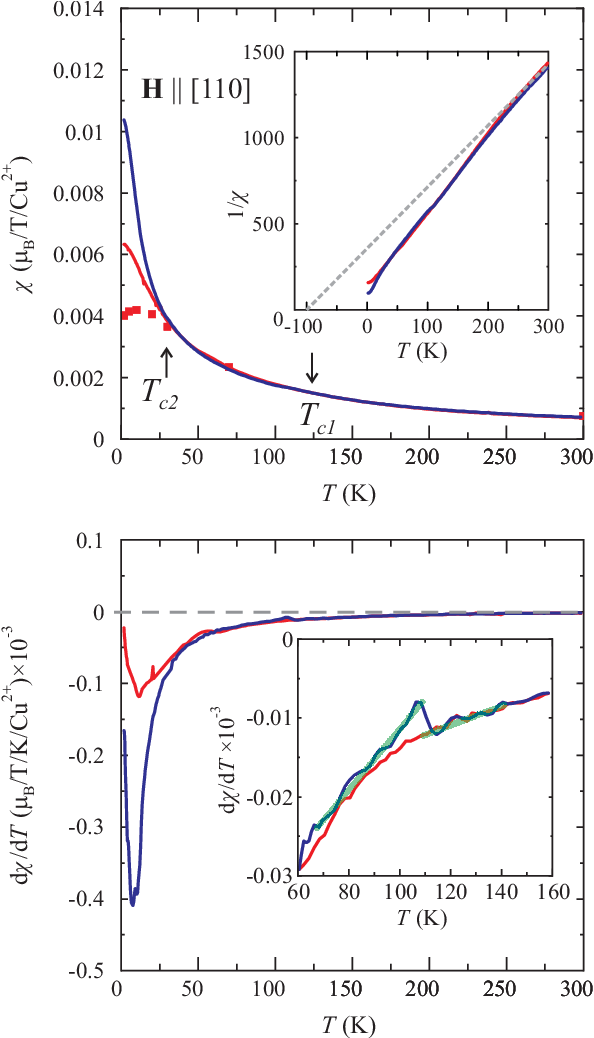}
\caption{(color online) Top panel: Temperature dependence of magnetic susceptibility $\chi(T)$ of \LiCu measured at $\mu_0H=1$~T (blue line) and $\mu_0H=5$~T (red line), $\vect{H}\perp C_4$. Red squares show temperature dependence of the value of differential susceptibility obtained from $M(H)$ measurements at $\mu_0 H=5$~T (see Fig.~\ref{fig7_M(H2)}). Inverse susceptibilities $1/\chi(T)$ are shown in the inset to the Figure 
with respective colors.
Bottom panel: Temperature derivatives of $\chi(T)$ from top panel. In the inset an expanded area in the vicinity of the singularity observed at $T\approx120$~K is presented. Green solid lines mark the smoothed slopes of $\chi(T)$ at both sides of the critical point.}
\label{fig4_M(T2)}
\end{figure}

Figs.~\ref{fig3_M(T1)} and~\ref{fig4_M(T2)} show the temperature dependences of magnetic susceptibility $\chi=M/\mu_0H$ measured at fields $\mu_0H$=1~T and $\mu_0H$=5~T at $\vect{H}$ directed along and perpendicular to C$_4$ axis of the crystals, respectively. Magnetization in all figures is given in units of $\mu_B$ per one magnetic ion of Cu$^{2+}$, assuming that our samples are stoichiometric. Growth of susceptibility with temperature decrease is observed in the whole temperature range. The dependences are well reproducible and no irrevercibility or history effects have been observed.
The temperature dependences can not be simply described by Curie-Weiss law: the reciprocal susceptibilities are not linear in studied temperature range (see insets to the top panels of Figs.~3,~4). The high temperature range can be roughly extrapolated by Curie-Weiss law with negative Curie-Weiss temperature $\Theta\approx$~-100~K, that allows to suppose the presence of strong dominant antiferromagnetic interaction in \LiCu. Against the background of susceptibility increase, a singularity at $T_{c1}=120\pm10$~K is observed. This singularity is accompanied by sharp change in slope of susceptibility. $d\chi(T)/dT$ dependences are shown in bottom panels of Figs.~\ref{fig3_M(T1)},~\ref{fig4_M(T2)}. The insets to the bottom panels of Figures show dependences $d\chi(T)/dT$ expanded in the vicinity of the singularity at T$_{c1}$. Green solid lines show the smoothed dependences above and below T$_{c1}$. In line with results of NMR experiments discussed below, we ascribe the singularity to partial antiferromagnetic ordering of \LiCu. In this case, at lower temperature the susceptibility of ordered part stops its growth with temperature decrease, whereas the rest of the magnetic system continues to behave paramagnetically.

The temperature dependences of susceptibilities measured at $\mu_0H=1$~T and $\mu_0H=5$~T are close at high temperatures, whereas at low temperatures they differ. This happens at temperatures below T$_{c2}\approx30$~K. This is the marker of nonlinearity of magnetic moment dependence on magnetic field at low temperatures.

\begin{figure}[tb]
\includegraphics[width=0.9\columnwidth]{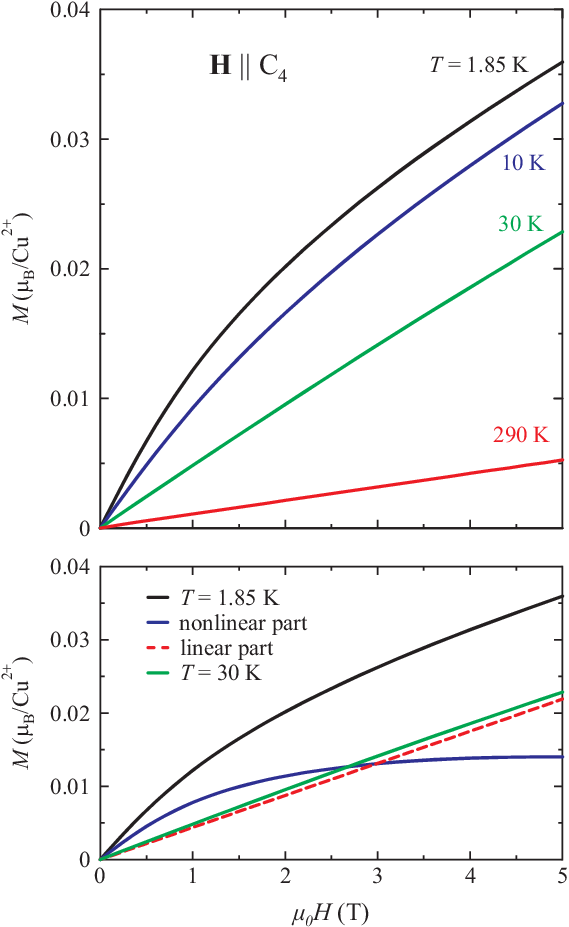}
\caption{Top panel: Field dependence of magnetization $M$ of \LiCu\ measured at different temperatures, $\vect{H}\parallel C_4$. 
Bottom panel: M(H) measured at $T=1.85$~K (black line) and at $T=30$~K (green line). Linear and nonlinear parts of {$M(H, T=1.85$~K$)$} obtained as described in the text are shown with dashed red and blue lines, respectively.}
\label{fig6_M(H1)}
\end{figure}

\begin{figure}[t]
\includegraphics[width=0.9\columnwidth]{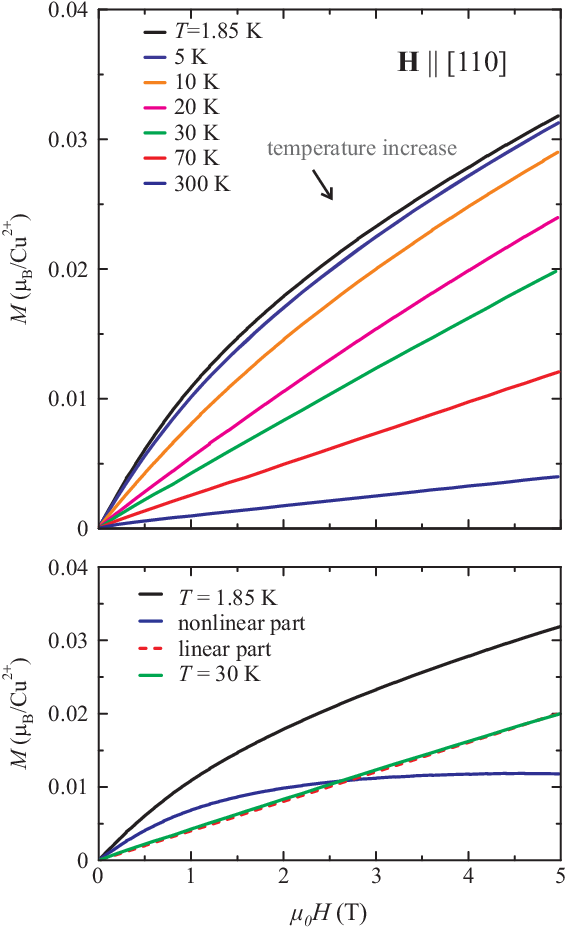}
\caption{Top panel: Field dependence of magnetization $M$ of \LiCu\ measured at different temperatures, $\vect{H}\parallel[1\overline{1}0](\perp C_4)$. 
Bottom panel: M(H) measured at $T=1.85$~K (black line) and at $T=30$~K (green line). Linear and nonlinear parts of {$M(H, T=1.85$~K$)$} obtained as described in the text are shown with dashed red and blue lines, respectively.}
\label{fig7_M(H2)}
\end{figure}

Field dependences of magnetic moment of \LiCu\ at different temperatures for field parallel and perpendicular C$_4$-axis are shown in upper panels of Figs.~\ref{fig6_M(H1)} and~\ref{fig7_M(H2)}. The dependences are linear at temperatures higher than $T\approx30$~K and nonlinear in low-temperature range. The differential susceptibilities $d\chi/dH$ at high fields ($\approx 5$~T)  for all curves measured below 30~K are nearly the same. The bottom panels of Figs.~\ref{fig6_M(H1)} and ~\ref{fig7_M(H2)} show the results of separation of $M(H)$ at $T=1.85$~K in a linear and a nonlinear parts: experimental black curve measured at 1.85~K can be presented as the sum of linear part (dashed red line) and nonlinear part (blue line). Linear part is close to $M(H)$ measured at 30~K. We ascribe the linear part of $M(H)$ below $T_{c2}$ to magnetization of antiferromagnetically ordered fraction, and nonlinear part -- to magnetization of paramagnetic fraction. Note, that linear part of susceptibility is weak: 
at $\mu_0 H=5$~T, the value of magnetic moment $M$ amounts $\approx2$~\%  of the saturated value $gS\mu_B/$Cu$^{2+}$.  
The maximum contribution of nonlinear part to full magnetization at the lowest temperature and $\mu_0 H=5$~T is $\delta M \approx 0.012-0.014$~$\mu_B/$Cu$^{2+}$ (see bottom panels in Figs.~\ref{fig6_M(H1)},~\ref{fig7_M(H2)}). 
This contribution can be reasonably explained by magnetization of the isolated magnetic ions discussed in chapter II (bold blue cross in Fig.~\ref{fig2_schema}). The estimated portion of isolated Cu$^{2+}$ ions is 1.6\%, which is close to experimental value. 
Note, that the similar nonlinear contribution to $M(H)$ was observed in weakly interacting antiferromagnetic nanoparticles~\cite{Balaev_2014}.

The red squares in dependences of $\chi$(T) in Figs.~\ref{fig3_M(T1)},~\ref{fig4_M(T2)}  show the values of differential susceptibility $dM/dH$ obtained from $M(H)$ measurements at $\mu_0 H=5$~T, which are close to the linear part of susceptibility presumably defined by ordered magnetic fraction of \LiCu. The temperature at which the differential susceptibility measured at $\mu_0 H= 5$~T stops increasing is signed as $T_{c2}\approx$30~K. Presumably, we associate this temperature with ordering temperature of residual portion of magnetic ions of \LiCu\ which were not ordered at $T<T_{c1}=123$~K.

\begin{figure}[t]
\includegraphics[width=1\columnwidth]{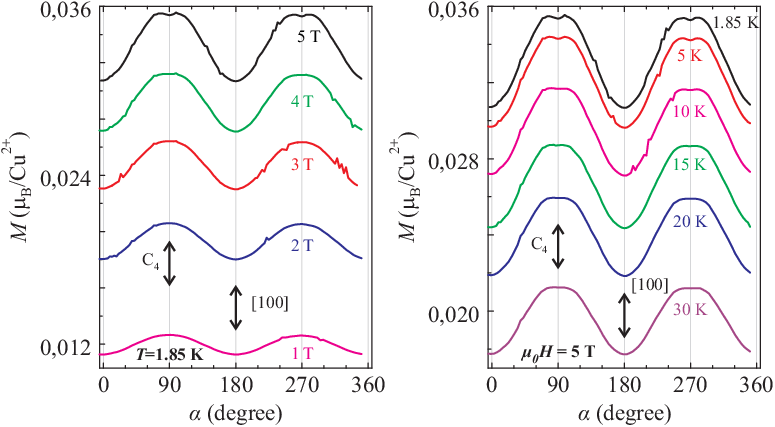}
\caption{Left panel: Angle dependence of magnetization $M(\alpha)$ of \LiCu\ measured at different values of $\mu_0H$,  $T=1.85$~K. 
Right panel: Angle dependence of magnetization $M(\alpha)$ of \LiCu\ measured at different temperatures T,  $\mu_0 H=5$~T.}
\label{fig8_M(a)}
\end{figure}

Left panel of Fig.~\ref{fig8_M(a)} shows angle dependences of magnetic moment $M(\alpha)$ of \LiCu\ measured at different values of magnetic fields at constant temperature $T=1.85$~K, where $\alpha$ is the angle between $[100]$-axis and field direction. Field was rotated from [100]-axis to $C_4$-axis of the crystal. Right panel of Fig.~\ref{fig8_M(a)} shows the temperature evolution of $M(\alpha)$, measured at $\mu_0H$=5~T. In whole studied field and temperature range susceptibility along $C_4$ exceeds susceptibility in plane (001) by $16\pm 1$\%. This value can be explained by anisotropy of gyromagnetic ratio of Cu$^{2+}$. The close value of anisotropy of $g$-factor was observed in related magnets LiCu$_2$O$_2$ and LiCuVO$_4$: $\Delta g/g\approx$ 10\% and 14\%, respectively \cite{Vorotynov_1998,buttgen_2007}.

	\begin{figure*}[t]\centering
\includegraphics[width=2\columnwidth]{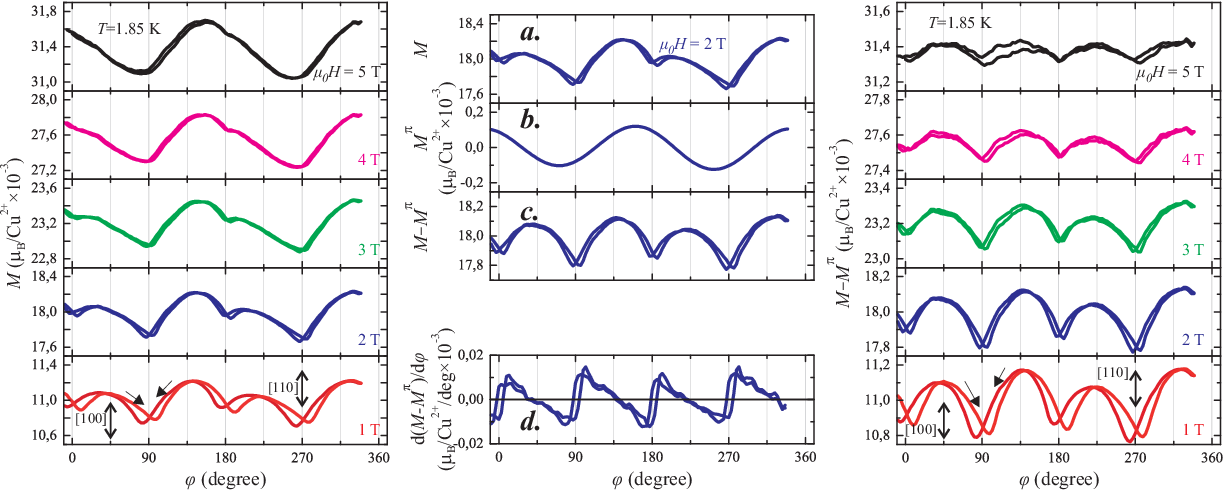}
\caption{Left panel: angle dependences of magnetic moment $M$ of \LiCu\ measured at $\mu_0 H=1-5$~T, $T=1.85$~K. Middle panel: {\bf{a}}. $M(\varphi)$ measured at $\mu_0 H=2$~T; {\bf{b}}. $\pi$-harmonic of $M(\varphi,\mu_0 H=2$~T$)$ harmonic series; {\bf{c}}. results of subtraction of $\pi$-harmonic from $M(\varphi,\mu_0 H=2~T)$; {\bf{d}}. angle derivative of $M(\varphi,\mu_0 H=2$~T$)-M^{\pi}(\varphi,\mu_0 H=2~T)$. Right panel: results of subtraction of 2nd harmonics $M^{\pi}(\varphi)$ from magnetization measurements $M(\varphi)$. Arrows at the curves in the figure show the directions of field rotations.}
\label{fig10_Minplane}
\end{figure*}

Left panel of Fig.~\ref{fig10_Minplane} shows the angular dependences of magnetization $M(\varphi)$ at $\vect{H}$ applied in plane perpendicular to $C_4$., $\varphi$ is the angle between $[110]$-axis and field direction. $M(\varphi)$ were measured at different fields, $T=1.85$~K. The angle dependent part of susceptibility at all studied fields and temperatures does not exceed 3\% of mean value. The angle dependent part can be divided in two components: one harmonic component with period $180^\circ$, and another component with period $90^\circ$ (see middle panel of Fig.~\ref{fig10_Minplane}, a-c). Taking into account crystal symmetry of \LiCu, the $\pi/2$-anisotropy seems to be natural. The harmonic modulation with period $180^\circ$ we presumably associate with parasitic signal from the sample holder. The amplitude of this modulation was proportional to applied field. Right panel of Fig.~\ref{fig10_Minplane} shows the dependences given in left panel of Fig.~\ref{fig10_Minplane} with subtracted $\pi$-harmonics.
Next, we list the main features of $90^\circ$-periodic parts of $M(\varphi)$. The maxima of magnetization is observed at $\vect{H}$ directed along square sides of crystal structure ($H\parallel$ $[1 0 0]$, $[0 1 0]$). Attention is drawn to the sharp change in the sign of the slope to the left and to the right of the minimum. In the vicinity of the minima, the dependences measured in clockwise direction differ from the curve measured in counter-clockwise rotation.  Arrows in the figure show the directions of field rotations. Slope singularity (see middle panel of Fig.~\ref{fig10_Minplane}d) and hysteresis mark the presence of the reorientation transitions at these angles. Such behavior is specific for exchange structures with axial symmetry such as collinear structure defined by vector of antiferromagnetism  $\bold{l}$, or planar spiral structure defined by vector $\bold{n}$ normal to spin plane; for order parameters $\bold{l}$ or $\bold{n}$  there are strong easy plane anisotropy within (001)-plane and weaker anisotropy within easy plane. The examples of collinear and spiral antiferromagnets with such anisotropy hierarchy are well known (see, for instance, Refs.~\cite{Glazkov_2023, Sakhratov_2018}).

\begin{figure}[t!]
\includegraphics[width=0.9\columnwidth]{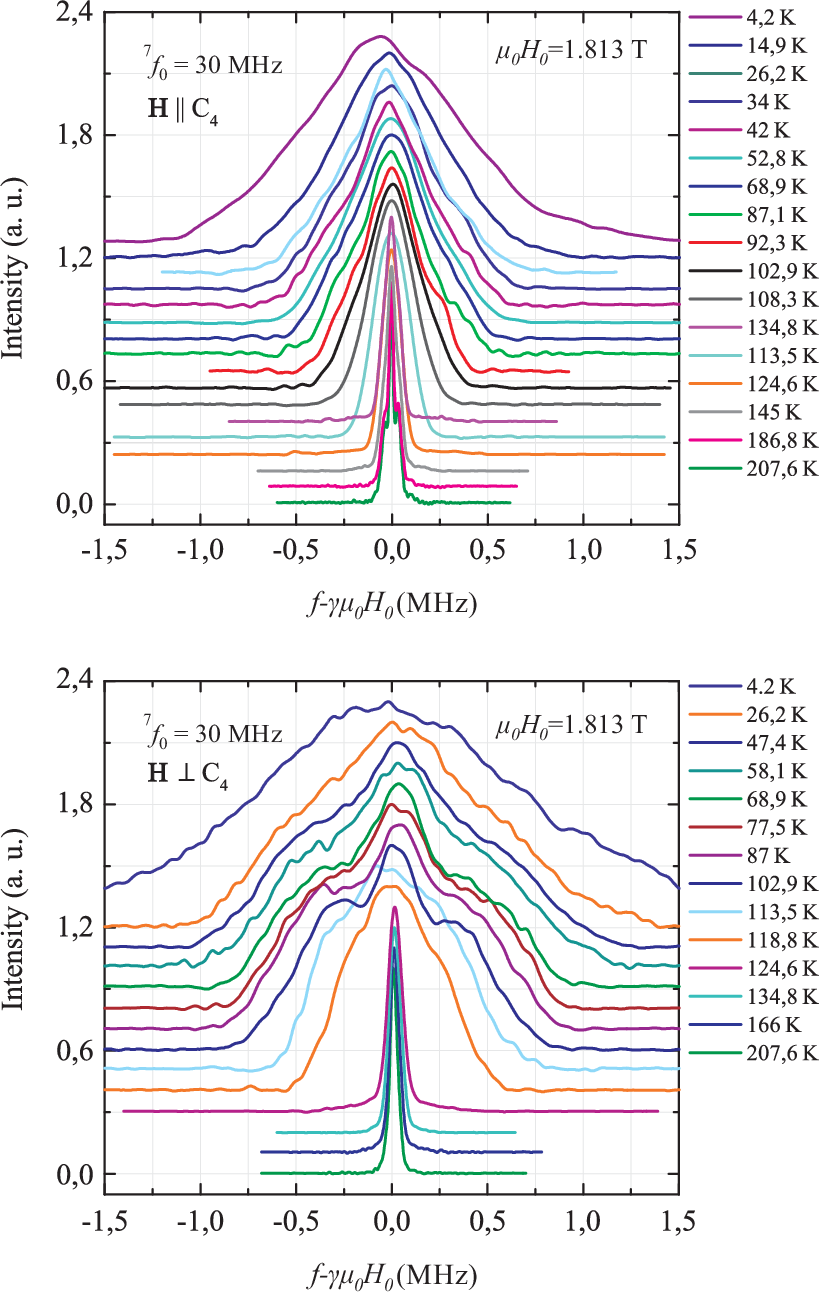}
\caption{Temperature evolution of $^{7}$Li NMR spectra at $\vect{H}$ parallel and perpendicular to the crystallographic axis $C_4$; $\mu_0 H_0=1.813$~T. NMR lines are stacked along ordinate axis for clarity.}
\label{fig11}
\end{figure}

 \begin{figure}[t]
\includegraphics[width=1\columnwidth]{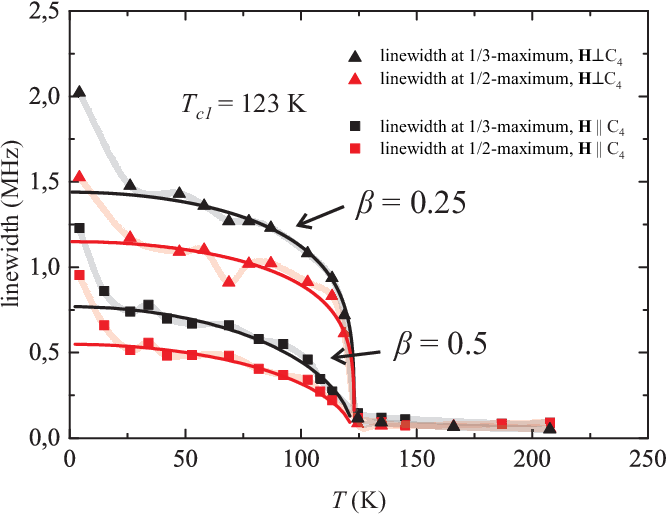}
\caption{Temperature dependence of $^7$Li NMR spectra linewidth $\delta \nu (T)$ measured at 1/2 (red symbols) and 1/3 (black symbols) of maximum. Squares correspond to $\vect{H}\parallel C_4$ and triangles correspond to $\vect{H} \perp C_4$. Transparent lines are given as guides for eye. Solid lines present the fit to the power law behavior with exponents $\beta=0.5$ for $\vect{H}\parallel C_4$ and $\beta=0.25$ for $\vect{H}\perp C_4$ and $T_{c1}=123$~K.} 
\label{linewidth}
\end{figure}

 \subsection{Nuclear magnetic resonance.}
The $^{7}$Li NMR spectra of \LiCu\ were measured at the frequency scan around $\nu_0=30$~MHz at permanent magnetic field equal to $\mu_0H_0$~=$\nu_0/\gamma=1.813$~T. Temperature evolutions of spectra for field parallel and perpendicular to the crystal axis $C_4$ are shown in Fig.~\ref{fig11}. At temperatures higher then $\approx 120$~K narrow NMR lines are observed. Lines obtained at H$\parallel C_4$ at $T=207$~K demonstrate the quadrupole splitting of central line of the $^{7}$Li NMR spectra corresponding to the central transition $m_I = -1/2 \leftrightarrow +1/2$ and two satellites corresponding to the transitions $m_I = \pm 3/2\leftrightarrow \pm 1/2$. The quadrupole splitting in \LiCu\ is weak and will not be considered in the following discussion of the spectra broadering observed at low temperatures. The observed lines demonstrate the strong change in linewidth and line shape at $T_{c1}\approx123$~K. The mass center of observed spectra does not shift distinctly with temperature, which is in agreement with the small susceptibility of \LiCu\ (see Figs.~\ref{fig6_M(H1)}--\ref{fig7_M(H2)}).
 
The temperature dependences of linewidths $\delta\nu(T)$  measured at the 1/2- and 1/3- level of the maximal NMR signal at $\vect{H}$ parallel and perpendicular to $C_4$ are shown in Fig.~\ref{linewidth}. Temperature increase of $\delta\nu(T)$ measured at 30~K~$<T<T_{c1}$ is proportional to $(1-T^2/T_{c1}^2)^\beta$ (see solid lines in the Figure). In our experiment, we obtain $\beta=0.50\pm 0.06$ for $\vect{H}\parallel C_4$ and $\beta=0.25\pm0.03$ for $\vect{H}\perp C_4$.
Experimentally observed disparity of exponents for different orientations of external magnetic field indicates that in-plane and out-of-plane components of magnetic moments  order by different scenarios. 

We associate the broadening of the NMR lines at $T< T_{c1}$ with the local magnetic fields from nearest ordered magnetic  Cu$^{2+}$ ions on the $^7$Li nuclei. 
The broad spectrum signals a continuous distribution of local magnetic field values or directions on $^7$Li positions with life-time higher than time of NMR measurement. 
Note, that broad continuous spectra are often found in incommensurate noncollinear spin structures~\cite{buttgen_2007,sakhratov_2014}.

Concluding, the abrupt broadenings of $^7$Li NMR line take place at $T_1\approx123$~K and $T_2\approx 30$~K which is in agreement with singularities on $M(T)$ observed at $T_{c1}$, $T_{c2}$.
The shape of the $^7$Li spectra indicates continuous distribution of values or directions of Cu$^{2+}$ magnetic moments in magnetically ordered state.

\section{Discussion}
We shall start the discussion from the model of infinite square lattice with sites occupied by magnetic ions of Cu$^{2+}$ (S=1/2) with the dominant in-plane exchange interactions along the sides and diagonals of squares J$_1$ and J$_2$, respectively,
hoping that this rough model can explain several observations.

Taking into account the results of magnetization measurements, we can consider that the dominant exchange interaction is antiferromagnetic, and as a result,  in the large-spin limit, we can expect the long range order to be antiferromagnetic. If the exchange interaction along square sides,  J$_1$, is dominant, the wave vector of the magnetic structure is $\bf{q}=[\pi, \pi]$. In case the dominant diagonal interaction is J$_2$, the structures with ${\bf{q}}=[\pi, 0]$ are expected. In the quantum case of $S=1/2$, corresponding short range correlations are expected. The model phase diagram for the large-spin limit and $S=1/2$ for 2D model was considered in Refs.~\cite{Chandra_1990, Valiulin_2019}.

\begin{figure}[t]
\includegraphics[width=0.9\columnwidth]{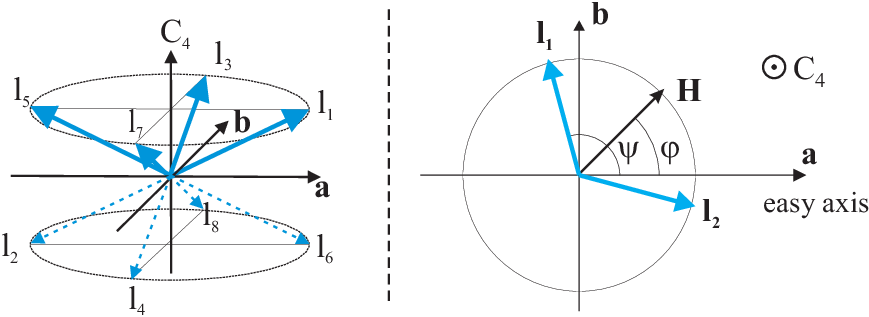}
\caption{Left panel: possible orientations of antiferromagnetic vector of model collinear structure at $H=0$. Right panel: Schema of a model two-domain structure with dominant easy-plane and weak in-plane anisotropies at field applied in easy plane. Angles are counted from the equilibrium direction of $\vect{l}$ at $H=0$. Here, $\vect{a},\vect{b}$ mark the easy axes in plane $(001)$.}
\label{domains}
\end{figure}

Possible orientations of antiferromagnetic vector $\bf{l}$ are defined by crystal anisotropy. As far as crystal structure of \LiCu\  is uniaxial, in general case, the zero field directions of $\bf{l}$ will be along one of four axes lying in planes $(1 0 0)$ and $(0 1 0)$ or $(1 1 0)$ and $(1 \overline{1} 0)$ of the crystal.
For both cases we can expect 8 domains with equal energy at zero external field. Vectors $\bf{l}_{1-8}$ are shown with solid and dashed blue lines in left panel of Fig.~\ref{domains}. 

In the special case of easy-axis anisotropy, we can expect only two domains with $\bf{l}$ directed along $[001]$ and $[00\overline{1}]$. The clear observation of 90$^{\circ}$-anisotropy of  magnetization measured in \LiCu\ within $(001)$-plane excludes this limit case.

Angle dependences of magnetic moment $M(\varphi)$ for low field $\vect{H}$ applied in $(001)$-plane can be understood in the model of strong easy plane anisotropy and weak in-plane anisotropy.

Anisotropic part of energy of the magnetic system can be written as $D\cos^2 2\varphi-\dfrac{1}{2}\chi_{\perp}H^2\sin^2(\varphi-\psi)$, where the first term is in-plane anisotropy and the second term is Zeeman energy. Here $D$ is an in-plane anisotropy constant and $\chi_{\perp}$ is the magnetic susceptibility at $\bf{H}$ applied perpendicular to $\bf{l}$. $\varphi$ and $\psi$ are the angles defining the directions of vectors $\bf{l}$ and $\bf{H}$, respectively, counted from equilibrium direction of $\bf{l}$ at $H=0$ (see right panel of Fig.~\ref{domains}). At field directed exactly between easy axes, domains $\pm l_1$ and $\pm l_2$ have equal energy. At small clockwise rotation of magnetic field, domains with $\pm\vect{l}_1$ will be preferable, whereas at rotation in other direction, domains with $\pm\vect{l}_2$ have lower energy. At this field direction, a spin-flop reorientation can be expected. The simple considerations of angle dependence of magnetic moment measured at different fields show that the smallest value of magnetization is expected at reorientation angle, whereas the largest one -- at $\bf{H}$ directed along easy axis. The amplitude of magnetization modulation at field rotation in low-field limit is expected to be close to 50\% of moment measured at $\vect{H}$ parallel to easy axis. 

The magnetization study of \LiCu\ revealed an anomaly at $\vect{H}\parallel[110]$ corresponding to spin-flop reorientation. Experimental observation of spin-flop reorientation naturally allows to suggest that the magnetic structure of \LiCu\ has a uniaxial anisotropy, 
and axes $[100]$ and $[010]$ of the crystal are easy directions for the specific axis of the magnetic structure. In the contrast to collinear model antiferromagnet discussed before, the maximal modulation observed in the experiment was approximately 3\% of maximal value. Such small value of modulation can be explained by considering the magnetic stucture of \LiCu\ as uniaxial with anisotropy of susceptibility $\chi_{\parallel}/\chi_{\perp}\approx~0.94^{\pm1}$. Here, indexes $\parallel$ and $\perp$ correspond to mutual orientation of field and the specific axis of the uniaxial magnetic structure. Note, that in case of 
a usual collinear antiferromagnet $\chi_{\parallel}/\chi_{\perp}\approx 0$ at low temperatures.

Next, we will suggest more realistic model of magnetic structure of \LiCu\ which can qualitatively explain some experimental observations, particularly, weak anisotropy of magnetic susceptibility. 
The magnetic ions in \LiCu\ have different environment of magnetic Cu$^{2+}$ and nonmagnetic Li$^+$. We can evaluate the number of magnetic ions which have all four bonds in square planes between nearest magnetic ions which are marked with red crosses in our random sample of A and B planes in Fig.~\ref{fig2_schema}. In the A-plane, we expect $32$\% of all sites to be occupied by such ions, whereas for the B-plane the ratio is even  smaller ($7.8$\%).
The number of such ions in both planes is much smaller than the percolation treshold for square lattice~\cite{Ziman}.
Thus, ions with full set of bonds with neighbors in planes A and B form clusters joint together with magnetic ions with incomplete set of bonds.  

For A-planes, clusters of ions with all four bonds are big and they all are connected via ``bridges'' forming an infinite cluster, because the concentration of magnetic ions in A-planes is higher than percolation threshold. 
For B-planes, clusters of ions with full set of bonds are small, and overall concentration of magnetic ions is close to the percolation threshold which allows to suggest that magnetic ions in B-planes form weakly coupled clusters of finite sizes.  
Particularly, cluster that consists of a single magnetic ion (Fig.~\ref{fig2_schema}b, blue cross), as discussed before, stipulates for a sufficient part of magnetic susceptibility at low magnetic fields at $T<30$~K.

Magnetic transition at $T_{c1}\approx$~123~K can be associated with the ordering in A-planes.
In the clusters with full set of bonds, one can expect the establishing of a collinear structure with antiferromagnetic vector $\vect{l}$ directed along one of the easy axes (as discussed before).  
``Bridges'' of magnetic structure that have not all 4 bonds possibly form a noncollinear structure either due to the frustration of magnetic bonds, or because they connect collinearly ordered areas with different orientations of $\vect{l}$.
Thus, magnetic state in A-planes can be concidered as a multidomain structure with strongly developed domain walls: the number of magnetic ions forming the domain walls 
 can be roughly evaluated as a number of ions with noncomplet number of bonds, i.e. $\approx 60$ \% of magnetic ions  in A-plane.
Application of a sufficiently strong magnetic field must decrease the number of possible directions of $\vect{l}$ in magnetic domains, although complete monodomenization of the magnetic structure is impossible because magnetic field interacts with domains with $\vect{l}$ and $-\vect{l}$ equally.

Low-temperature transition at $T_{c2}\approx 30$~K supposedly correspond to magnetic ordering in B-planes. 
Note that the expected value of magnetic ions in these planes with full set of bonds with magnetic neighbors is equal to $0.6^5\cdot 100\% =7.8\%$.
Proximity of magnetic ions concentration in B-planes to percolation threshold allows to suggest that connections between clusters are weak. Magnetic order in clusters of different sizes can occur at different temperatures, so the magnetic ordering takes place at a broad temperature range, which is the reason it does not manifest itself as an abrupt singularity on temperature dependences of magnetization and NMR linewidth.
It is also possible that the ordering in B-planes has imposed character due to the interaction with ordered neighboring A-planes. 

\begin{figure}[t]
\includegraphics[width=0.9\columnwidth]{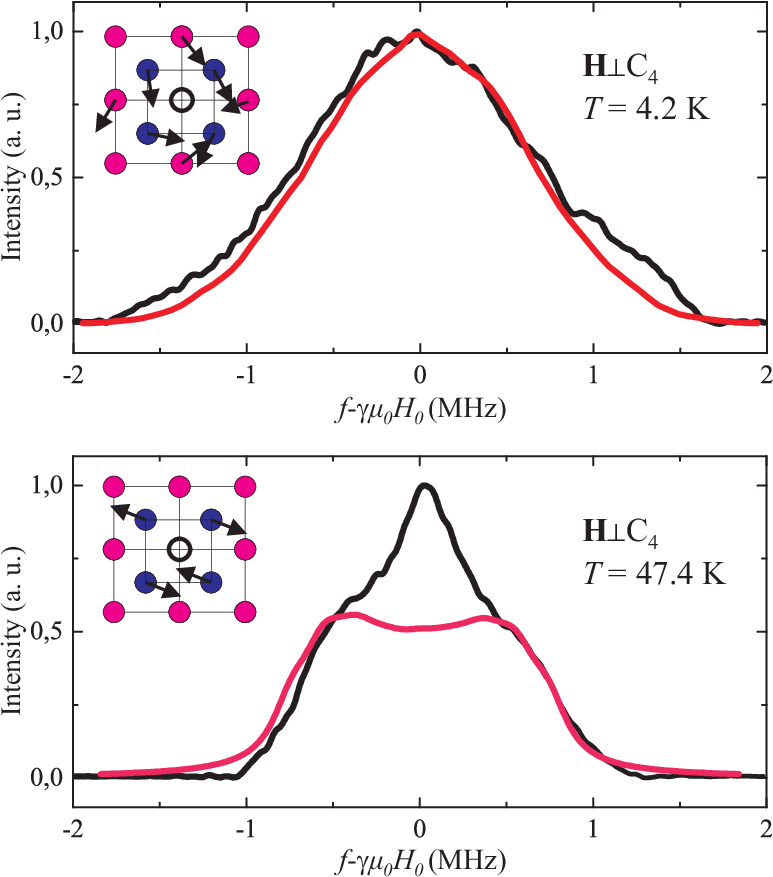}
\caption{The results of modelling $^{7}$Li NMR spectra, $\vect{H}\perp C_4$ from Fig.~\ref{fig11}. Black lines -- experimental spectra obtained at $T=4.2$~K (top panel) and $T=47.4$~K (bottom panel). Corresponding red lines -- spectra calculated in frames of the models described in text. Insets show the model orientation of magnetic moments  in crystallographic planes A (blue circles) and B (magenta circles).}
\label{NMR_theory}
\end{figure}

$^7$Li NMR spectra demonstrate drastic line broadening at $T<T_{c1}$, which can be ascribed to occurrence of static magnetic moments on Cu$^{2+}$ ions. These moments create an effective field $\Delta \vect{h}$ at $^7$Li nuclei which leads to the shift of NMR frequency on $\gamma \Delta h_H$. Here, $\Delta h_H$ is a projection of local field $\Delta\vect{h}$ on direction of external magnetic field $\vect{H_0}$.

For modeling of NMR spectra, only dipole fields from Cu$^{2+}$ magnetic ions from the first coordination spheres of Li$^{+}$ ions were considered. Such a suggestion was previously justified for related noncollinear structures in LiCuVO$_4$ and RbFeMoO$_4$~\cite{Buttgen_2010,Soldatov_2020}. We also assumed that the values of all Cu$^{2+}$ magnetic moments were the same. 

Two different model structures were considered: the model with the ordered moments directions randomly distributed within the square plane perpendicular to $C_4$-axis and the model with ordered moments directions randomly distributed in space.
$^7$Li NMR spectra calculated in frames of listed suggestions are presented in upper panel of Fig.~\ref{NMR_theory} with red line. They are in agreement with spectrum obtained experimentally at lowest temperature $T=4.2$~K shown in the same Figure with black line.
The shapes of calculated spectra were the same for both models with  values of Cu$^{2+}$ magnetic moments equal to 0.4~$\mu_B$ for planar structure and 0.6~$\mu_B$ for the model of moments randomly directed in space. The width of individual NMR lines used for computations was $\delta\nu=0.045$~MHz.
In the inset to the Figure~\ref{NMR_theory}, the schema of nearest to Li$^{+}$ randomly directed moments of Cu$^{2+}$ is presented. Li$^{+}$ is marked with open circle. Square lattice sites are marked with blue for ions from  A-plane and red for ions from B-plane.
In the frames of these models, neighboring positions of  Li-ions were randomly occupied with magnetic Cu$^{2+}$ and non-magnetic Li$^{+}$ ions in accordance with occupation probability in A- and B-planes.
Strong spin reduction in low-dimensional $S=1/2$ systems is natural: for related magnets Sr$_2$CuO$_3$ and LiCuVO$_4$, magnetic moment value is 0.06~$\mu_B$ and 0.31~$\mu_B$, respectively~\cite{Kojima_1997,Gibson_2004}.

At higher temperatures, NMR lines have more complicated shape. 
Experimental spectrum obtained at $T=47.4$~K is shown with black line in the bottom panel of Fig.~\ref{NMR_theory} and model spectrum is shown with red line. The value of magnetic moment used in calculations was 0.4~$\mu_B$. The schema of suggested magnetic structure is given in the inset to Figure~\ref{NMR_theory}. At this temperature, we suppose that spins from B-plane are not ordered. In A-plane, the directions of closest magnetic moments were suggested to be correlated antiparallel and lying within the plane perpendicular to $C_4$, but their directions within the plane were arranged randomly.  This model partially describes the experimental spectrum at this temperature. Non-shifted peak can probably be explained with the signal from Li$^+$ surrounded with strongly reduced magnetic moments. The integral intensity of this peak is about $15\%$ of total NMR line intensity.

Here, we did not consider the possibility of the frustration of exchange interactions, which can play a significant role in formation of a frozen disordered structure. Hence, on the example of frustrated 2DQHAF square lattice system Li$_2$VOSiO$_4$ it was shown that doping with nonmagnetic Ti$^{4+}$ impurities sufficiently influences frustration of exchange interactions~\cite{Pappinutto_2005}.
In geometrically frustrated quasi-2D magnets with triangular lattice one can expect an establishing of a disordered spin-glass-like state even at 10-20\% depletion of the magnetic system with non-magnetic ions~\cite{Yan_2016}. Studies of magnetic properties of CuCrO$_2$, an example of quasi-2D antiferromagnets with triangular lattice, doped with vanadium -- CuCr$_{1-x}$V$_x$O$_2$  -- revealed spin-glass state at doping level ~8\%~\cite{Singh_2012, Kumar_2013}. Studies of another quasi-2D magnet, Rb$_{1-x}$K$_x$Fe(MoO$4$)$_2$, showed that samples with doping of 15\% ($x=0.15$) present a frozen magnetic disorder~\cite{Sakhratov_2019}.

\section{Conclusion}

In this work, a novel quasi-2D antiferromagnet with 
randomly depleted square lattice of spins $S=1/2$ was studied. 
$^7$Li NMR and magnetization measurements performed on single crystals of \LiCu\ revealed the occurrence of magnetic order at $T_{c1}=123$~K and the change of the magnetic state at $T_{c2}\approx30$~K. We ascribe the high temperature transition to the establishment of magnetic order in planes with higher concentration of magnetic ions and low temperature transition -- to the magnetic ordering in planes with lower concentration of magnetic ions. Broad continuous NMR spectra below $T_{c1}$ reflect a continuous distribution of magnetic moments in \LiCu , which is typical for spiral, spin-modulated magnetic structures or structures with frozen disorder.
Magnetization measurements revealed a spin-flop transition which idicates weak uniaxial anisotropy of spin structure with susceptibilities $\chi_{\parallel}/\chi_{\perp}\approx0.94$. Fourth-order anisotropy with easy axes directed along [100] and [010] axes of the crystal was observed. Small value of magnetic susceptibility at all orientations of applied magnetic field shows that magnetic structure is rigid,
since the estimated value of saturation field derived from differential susceptibility measured at $\mu_0 H=5$~T  is $\mu_0H_{sat} \approx$200~T.

\acknowledgments
We thank V.~I.~Marchenko, S.~S.~Sosin and M.~Zhitomirsky for stimulating discussions and L. B. Lugansky for data processing assistance.

The work was financially supported by Russian Science Foundation, Grant No 22-12-00259 (NMR data processing and calculations); Basic research program of HSE University (magnetization data processing); part of the work (on single crystals growth) performed at the MIREA RTU was funded by the Ministry of Science and Higher Education of the Russian Federation within the framework of the State Task (FSFZ-2023-0005 project).

\end{document}